\documentclass{csfourzero}

\title{How is the speed of code review affected by activity, usage and code quality?}
\author{William Brown}
\date{\today}
\usepackage{url}
\usepackage{natbib}
\usepackage{float}
\usepackage{color, graphicx, titlepic, pdflscape, lipsum, caption, tabularx, appendix}
\usepackage{subcaption}
\usepackage{hyperref}

\usepackage{amsmath,amssymb,amsthm}
\usepackage{thmtools}
\declaretheoremstyle[
spaceabove=6pt, spacebelow=6pt,
headfont=\normalfont\bfseries,
notefont=\mdseries, notebraces={(}{)},
bodyfont=\normalfont,
postheadspace=0.6em,
headpunct=:
]{mystyle}
\declaretheorem[style=mystyle, name=Hypothesis, preheadhook={}]{hyp}

\usepackage{cleveref}
\crefname{hyp}{hypothesis}{hypotheses}
\Crefname{hyp}{Hypothesis}{Hypotheses}

\defcitealias{TestCoverageMediaWiki}{MediaWiki test coverage}
\defcitealias{ExtensionTestCoverageMediaWiki}{MediaWiki extension test coverage}
\defcitealias{mediawiki:GitReviewers}{Git Reviewers list}
\defcitealias{mediawiki:developersmaintainers}{Code maintainers list}
\defcitealias{WikiApiaryAbout}{WikiApiary}
\defcitealias{WikiApiaryExtensions}{WikiApiary}
\defcitealias{GerritEfficiency}{Git Efficiency}
\defcitealias{PhabricatorBacklog}{Phabricator Backlog}
\defcitealias{GitOverview}{Git Overview}
\defcitealias{GerritApprovals}{Gerrit Approvals}
\defcitealias{BitergAbout}{Wikimedia Bitergia Analytics}
\defcitealias{GitExtensionsList}{list of extensions on MediaWiki's git repositories}

\usepackage[indent=0px, skip=10pt plus1pt]{parskip}
\PassOptionsToPackage{hyphens}{url}
\usepackage{etoolbox}
\AtBeginEnvironment{quote}{\par\singlespacing\small}

\abstract{This paper investigates how the speed of code review is affected by the code quality, activity and usage in the context of MediaWiki extensions. The median time to merge is compared against several other variables which are collected using a variety of manual methods and APIs. The results are graphed where possible and statistical analysis is used to determine the significance of the results. The paper finds that the number of reviewers voting on code and whether the extension has a steward affects the median time to merge. Finally, conclusions are drawn and further research topics are recommended.}

\begin{document}
\maketitle
\section{Introduction}

\label{sec:intro}

Code review is a vital part of the process of making changes to code, and slow code review creates a bottleneck that negatively affects the development and maintenance process of a project.

While code review is not the only possible bottleneck, in open-source projects those with the rights to merge are naturally a smaller group than those who can report bugs and write code to be submitted for review. This is because reliable code review is important and untrustworthy or inexperienced code reviewers could lead to a larger number of avoidable bugs and potentially even security issues. This means that code reviewers have to be vetted and inactive reviewers need to have their rights removed to avoid inactive accounts being compromised.

This means that keeping code review fast for the reviewers decreases the time they have to spend on each patch and with the same amount of time can review more patches. A too-slow code review process is likely to make volunteer developers and reviewers disinterested, which only makes it slower and could spiral to the project into inactivity.

\subsection{MediaWiki}

This project focuses on the open-source project MediaWiki, which is a project primarily written in PHP that is used to run Wikipedia along with its sister wikis \citep{WhatIsMediaWiki}. The base installation (named `core`) is all that is needed to run a wiki, however, installations can choose to extend the functionality of the wiki by installing extensions. These extensions are varied in terms of intended use, number of installations, code quality and maintenance levels \citep{mediawiki:manualextensions}.

The author of this paper chose MediaWiki as he has familiarity with writing and reviewing code for the project. He also chose it because of the large number of extensions which means a large pool of groupable data from the same project. This reduces the possibility that other factors could be affecting the process, as all extensions using a MediaWiki repository use the same code review and ticket reporting system.

Development of MediaWiki is supported and overseen by the Wikimedia Foundation (WMF) which also own the servers that host Wikipedia along with trademarks associated with Wikipedia \citep{WikiMediaFoundationAbout}. Different parts of the core installation and each extension used on sites run by the WMF will generally have a `steward` (usually a WMF team) along with maintainers listed on a centralised table \citep{mediawiki:developersmaintainers}.

One extension that the author is a reviewer and developer for is called CheckUser. This extension is used to combat the use of multiple accounts, but due to the sensitivity of the data it allows trusted users to see (IP addresses and UA strings of registered users) the code review process is more rigorous and often seems slower \citep{ExtensionCheckUser} \citep[p. 158]{10.1145/2030376.2030394}.

MediaWiki uses crowd-sourced translations from translatewiki.net for strings used as part of the interface \citep[p. 80]{fontanills2012panorama}. To achieve this, an automated bot creates and automatically merges commits that update the translations, which can also include back-porting the changes to previous release versions. Bots also are used to update libraries by approving a library upgrade once for all extensions and then the bot makes separate changes for each extension that uses the libraries.

\section{Background and Related Work}
\label{sec:lit}

Code review is important to assure quality of code \citep[p. 2183]{McIntosh2015}, however, slow code review `increases the risk of software degradation` \citep[p. 519]{9240657}. As such ensuring that code review is kept quick improves code quality, which means finding ways to speed up code review is important. Having efficient code review would also help to mitigate the impact of vulnerabilities \citep[p. 418]{10.1145/3379597.3387465}. Not having fast code review is considered bad practice by Google and Microsoft \citep[p. 7-8]{DOGAN2022106737}.

Without efficient code review reviewers may be put off reviewing patches, as research found that a reviewer will review if `benefit of review is higher than the cost` \citep[p. 67]{7809488}. While time it takes to review is not the only possible cost associated with a review, ensuring faster review will reduce the cost.

Research has found that tests have a `weak positive relationship` to bugs \citep[p. 112]{6605914}. This means that a reviewer is likely find review faster and easier if tests cover the changes, as it reduces the risk of a bug unseen by the reviewer causing issues.

Some MediaWiki extensions are more critical and sensitive than others which means that the depth of the code review may vary extension by extension. For example, research into how to perform `link spam` details how extensions like CheckUser are good at stopping the abuse \citep[p. 158]{10.1145/2030376.2030394}. An extension that has little usage and importance may have a less strict code review. Also, extensions that are deployed on WMF wikis generally have changes live on WMF wikis (including Wikipedia) a week after merge \citep{wikitech:Deployments/Train}. This means WMF deployed extensions are often subject to stricter code review standards.

A paper discussing lessons learned from writing software used by NASA for missions on Mars found that open-source projects outperformed paid-for solutions in terms of maintenance, code quality and support \citep[p. 44]{NorrisMissionCritical}. The bottleneck of slow code review is likely to slow progress and could lead to disinterest from both those writing code and reviewing code, so speedy code review is important to keep vital open-source projects running.

Finally, improvements in the review process were noted when reviewers and developers were automatically recommended for submitted changes \citep[p. 542]{7328331}. This also encourages more developers to also become reviewers \citep[pp. 507―508]{9240650}. MediaWiki allows users to be added to submitted changes using the \citetalias{mediawiki:GitReviewers}, however, an automatic system may improve speed of code review.

\section{Research Question}
\label{sec:rq}

\begin{quote}
    \begin{center}\Large{\textit{How is the speed of code review affected by activity, usage and code quality?}}\end{center}
\end{quote}

The `speed of code review` will be measured by the median time for a merge as defined by `Median time for a merge` section of the \citetalias{GerritEfficiency} panel on \citetalias{BitergAbout}. This shall be known as the target variable.

\label{list:activity-usage-and-code-quality}The `activity, usage and code quality` is measured in this paper using the following variables which could also be used to predict the target variable.
\begin{enumerate}
    \item \label{list:item-1}The PHP test case coverage of the code as measured by \citetalias{TestCoverageMediaWiki}
    \item \label{list:item-2}Whether the extension has a steward listed at the \citetalias{mediawiki:developersmaintainers}
    \item \label{list:item-3}Whether the extension has at least one maintainer listed in the \citetalias{mediawiki:developersmaintainers}
    \item \label{list:item-4}The number of changes submitted for review
    \item \label{list:item-5}The number of approved changes
    \item \label{list:item-6}The number of people who have authored approved changes
    \item \label{list:item-7}The number of code review votes
    \item \label{list:item-8}The number of reviewers voting on code
    \item \label{list:item-9}The average age for currently open tickets
    \item \label{list:item-10}Usage of the extension on all indexed wikis as reported by \citetalias{WikiApiaryAbout}
\end{enumerate}

Where relevant all bots are excluded from the statistics collected. This is because changes made or approved by bots are designed to be uncontroversial and as such are often merged without human review. Including them would skew the results and the conclusions drawn from the results would be less applicable to improving code review for human submitted patches.

\section{Experimental Design}
\label{sec:exp}

\subsection{Hypotheses}
Each of the 10 variables (excluding the target variable) listed in the \hyperref[list:activity-usage-and-code-quality]{Research Question} section have their own set of hypotheses. These hypotheses follow the form:
\setcounter{hyp}{-1}
\begin{hyp}[Null Hypothesis]\label{hyp:hyp-a}\textit{Variable} has no \textit{correlation to / effect on} the target variable.\end{hyp}
\begin{hyp}[Alternative Hypothesis]\label{hyp:hyp-b}\textit{Variable} has \textit{a correlation to / an effect on} the target variable.\end{hyp}

Where `\textit{Variable}` is replaced with one of the 10 variables and one of \textit{correlation} or \textit{effect} is chosen depending on if the data is graphed or tabled respectively. All but \hyperref[list:item-2]{variables numbered 2 and 3} will be graphed.

When plotted, the Pearsonr correlation coefficient of the data graphed and an associated p-value will be calculated. The correlation coefficient will be used to determine how linearly correlated the data is, where values closer to 1 or -1 are more linearly correlated. A high p-value (\(p > 0.05\)) suggests that the data is uncorrelated and as such \(H_0\) should be accepted. A low p-value (\(p < 0.05\)) and a Pearsonr far enough from 0 would indicate that \(H_1\) should be accepted \citep{scipy:pearsonr}.

For the data which is not graphed the two groups will be `Yes` or `No`. As such a t-test can be used to determine the significance of the differences between the values of the target variable in each group. This test returns the t-statistic and also an associated p-value that indicates how likely it would be to get this or a bigger t-statistic `from populations with the same population means` \citep{scipy:ttest}. If the p-value is too high then it suggests that \(H_0\) should be accepted. If the p-value is low enough and the t-statistic is far enough from 0 the results would then suggest \(H_1\) should be accepted.

\subsection{Dataset}
The dataset is created using data collected from each extension about each of the 10 variables from the \hyperref[list:activity-usage-and-code-quality]{Research Question} section. The list of extensions used is from the \citetalias{GitExtensionsList}. These extensions will have data about at least the variables \hyperref[list:item-4]{numbered 4 to 8}. If an extension does not have data on one of the other variables it is ignored when generating the results.

Data is manually parsed into a form that can be used by Python from \citetalias{ExtensionTestCoverageMediaWiki} for \hyperref[list:item-1]{variable number 1}. For \hyperref[list:item-2]{variables numbered 2 and 3}, the \citetalias{mediawiki:developersmaintainers} will be manually parsed into a form that can be used by Python. The \hyperref[list:item-10]{variable numbered 10} will be measured by downloading CSV (comma-separated values) files from \citetalias{WikiApiaryExtensions} and storing the number of sites using the extension.

Variables \hyperref[list:item-4]{numbered 4 to 9} will be measured using elastic searches to the APIs provided by \citetalias{BitergAbout} using elastic searches loosely based on selected queries from the \citetalias{GerritEfficiency}, \citetalias{PhabricatorBacklog}, \citetalias{GitOverview} and \citetalias{GerritApprovals} panels. This data will be collected such that results are collected from the last 9 months and also for all data (i.e. no time limit). This is to allow the author to later determine which is better when discussing the results. Because some extensions will have had no changes for the last 9 months, the 9-month dataset will naturally be smaller. 9 months was chosen as a good trade-off between getting as much data as possible while keeping the data from being too old.

\hyperref[list:item-1]{Variables 1} and \hyperref[list:item-10]{10} only holds data about the current state of the extension, and as such comparing it to a target variable generated without a time cutoff (potentially around 20 years worth of data for the oldest repositories) is unlikely to yield useful results. As such for these variables, the target variable is collected from data only over the last 9 months. While \hyperref[list:item-2]{variables 2 and 3} are similar to \hyperref[list:item-1]{1 and 10}, this list may not be fully up to date so data is collected to keep options open for the results.

\subsection{Dependence and independence}

Whether or not the variables being measured here are dependent or independent is not fully clear. The hypotheses attempt to determine if correlation could exist between the target variable and all other variables, and as such whether they are dependent is not known.

While dependency of variables being measured has not been established, the author notes that he thinks the number of code review votes is somewhat dependent on the number of submitted changes. Further investigation would be needed to establish dependence and independence of the variables.

\section{Results}
\label{sec:results}

\begin{figure}[H]
    \centering
    \includegraphics[width=211pt]{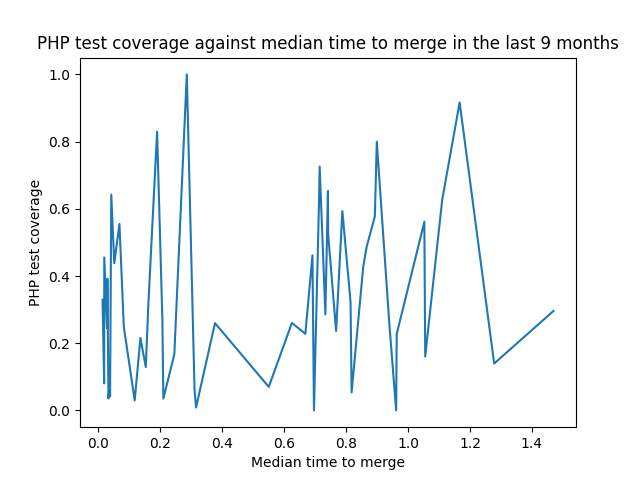}
    \caption{\hyperref[list:item-1]{PHP test coverage percentage}}
    \label{fig:number-of-changesets-9-m}
\end{figure}

These results cover the 911 extensions that are hosted on MediaWiki's git repositories. All results use all these extensions except for:
\begin{itemize}
    \item\hyperref[fig:number-of-changesets-9-m]{PHP test coverage} which only has data on 144 extensions as only these extensions had testing coverage reported
    \item\hyperref[sec:steward-team]{Whether an extension has a steward} which only has data on 171 extensions, as extensions can only have a steward if they are deployed on a WMF wiki \citep{mediawiki:developersmaintainers}.
    \item\hyperref[sec:listed-maintainers]{Whether an extension has at least one listed maintainer} which only has data on 170 extensions, as extensions only deployed on WMF wikis can have listed maintainers \citep{mediawiki:developersmaintainers}.
\end{itemize}

\subsection{Graphed results}

\begin{figure}[H]%
    \centering
    \subfloat[\centering Last 9 months]{{\includegraphics[height=5cm]{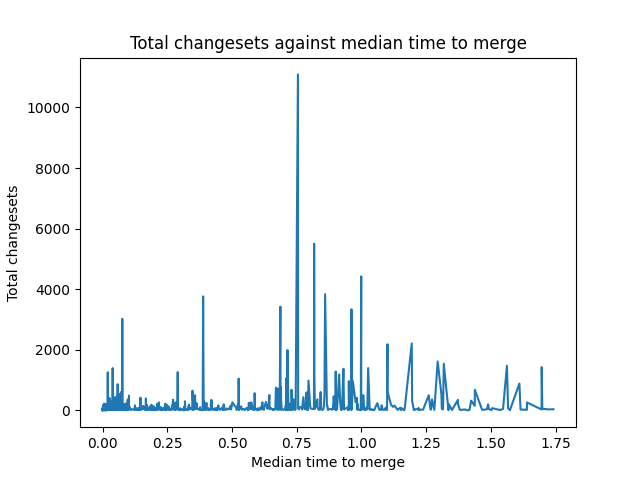} }}%
    \subfloat[\centering All data]{{\includegraphics[height=5cm]{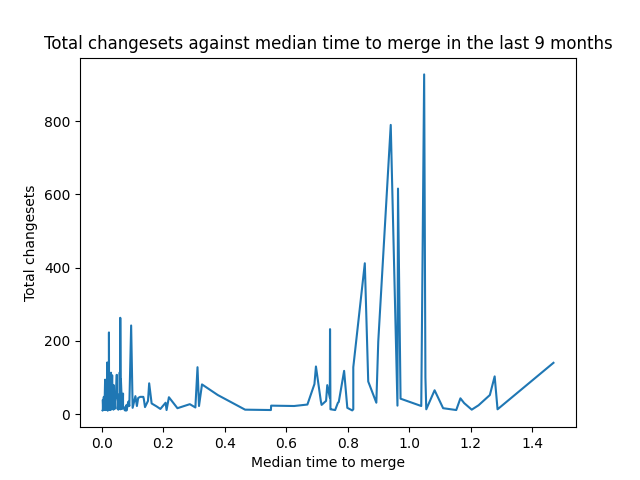} }}%
    \caption{\hyperref[list:item-4]{Number of submitted changes}}%
    \label{fig:number-of-changesets}%
\end{figure}

\begin{figure}[H]%
    \centering
    \subfloat[\centering Last 9 months]{{\includegraphics[height=5cm]{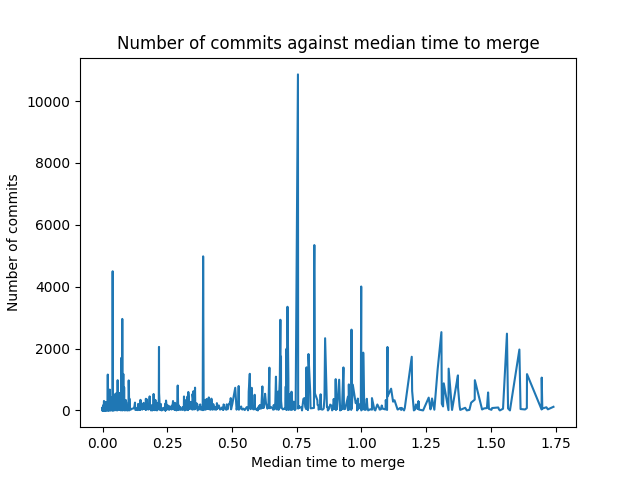} }}%
    \subfloat[\centering All data]{{\includegraphics[height=5cm]{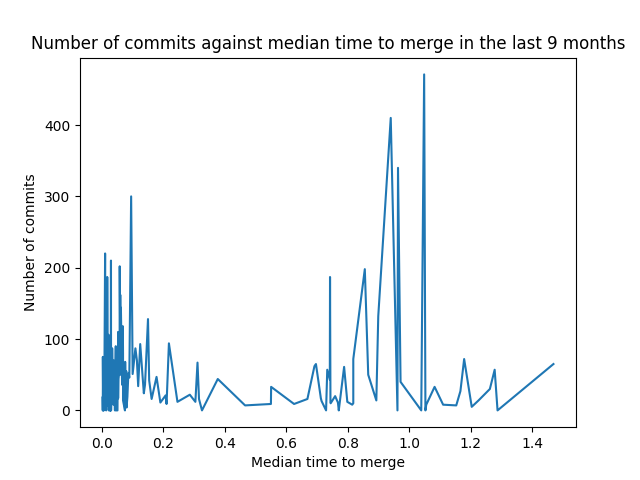} }}%
    \caption{\hyperref[list:item-5]{Number of approved changes}}%
    \label{fig:number-of-commits}%
\end{figure}

\begin{figure}[H]%
    \centering
    \subfloat[\centering Last 9 months]{{\includegraphics[height=5cm]{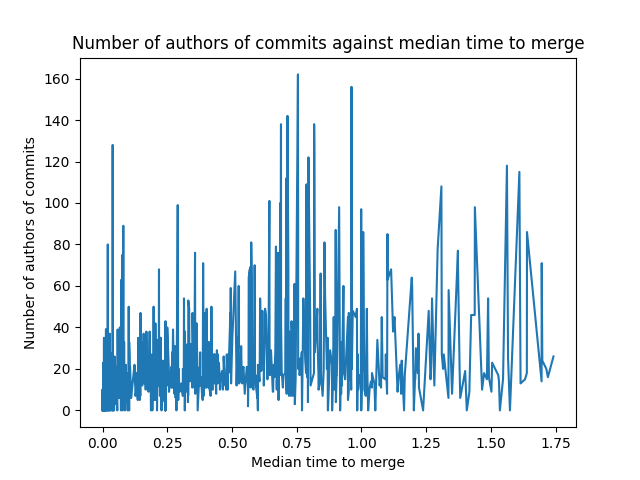} }}%
    \subfloat[\centering All data]{{\includegraphics[height=5cm]{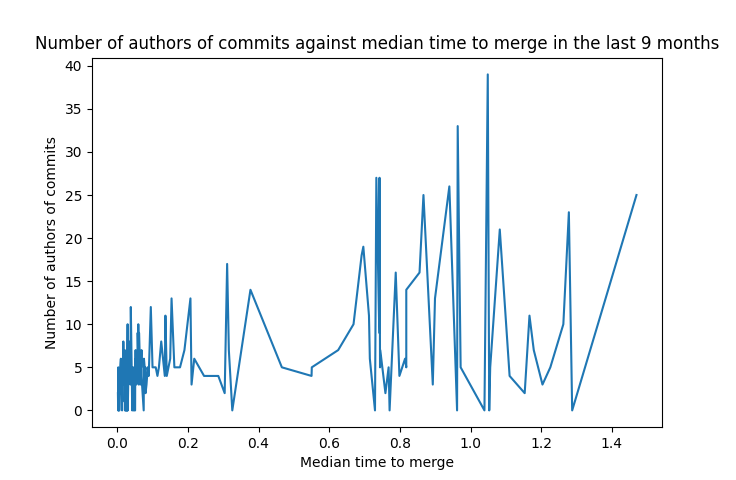} }}%
    \caption{\hyperref[list:item-6]{Number of authors who had changes approved}}%
    \label{fig:number-of-authors-committing}%
\end{figure}

\begin{figure}[H]%
    \centering
    \subfloat[\centering Last 9 months]{{\includegraphics[height=5cm]{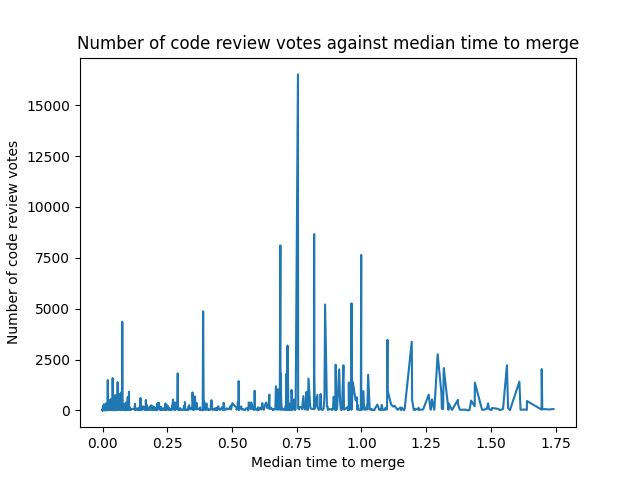} }}%
    \subfloat[\centering All data]{{\includegraphics[height=5cm]{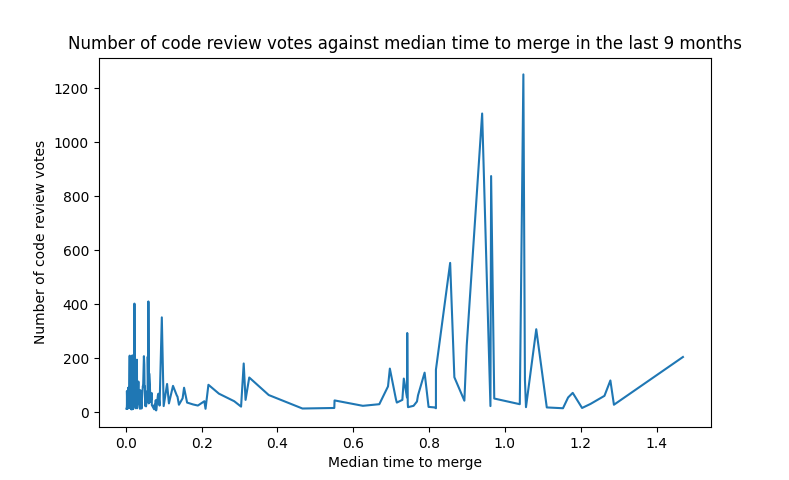} }}%
    \caption{\hyperref[list:item-7]{Number of code review votes on changes}}%
    \label{fig:number-of-code-review-votes}%
\end{figure}

\begin{figure}[H]%
    \centering
    \subfloat[\centering Last 9 months]{{\includegraphics[height=5cm]{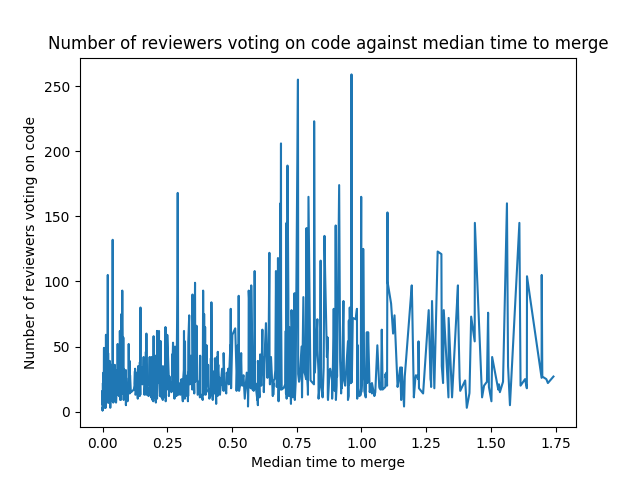} }}%
    \subfloat[\centering All data]{{\includegraphics[height=5cm]{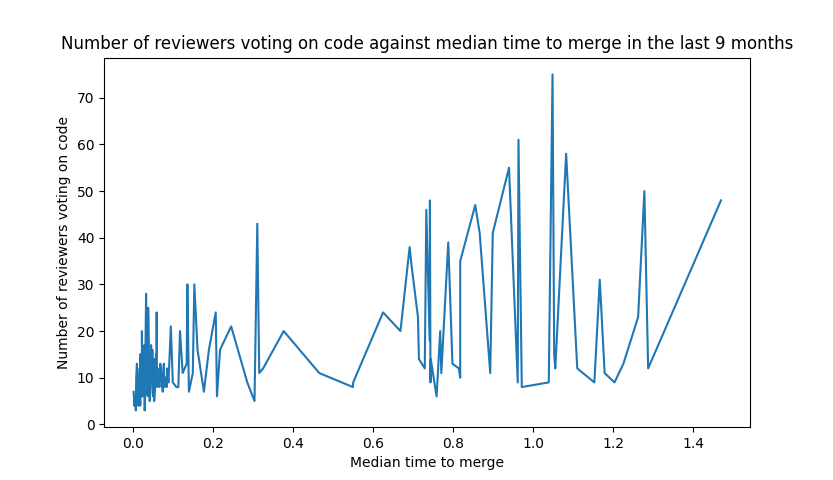} }}%
    \caption{\hyperref[list:item-8]{Number of reviewers who have voted on changes}}%
    \label{fig:number-of-reviewers-who-have-voted}%
\end{figure}

\begin{figure}[H]%
    \centering
    \subfloat[\centering Last 9 months]{{\includegraphics[height=5cm]{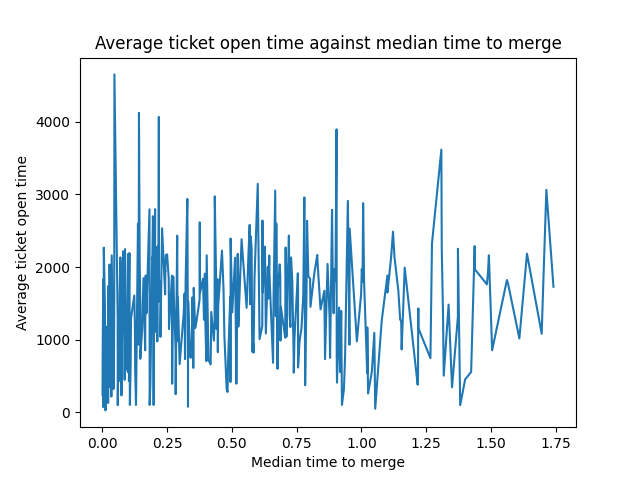} }}%
    \subfloat[\centering All data]{{\includegraphics[height=5cm]{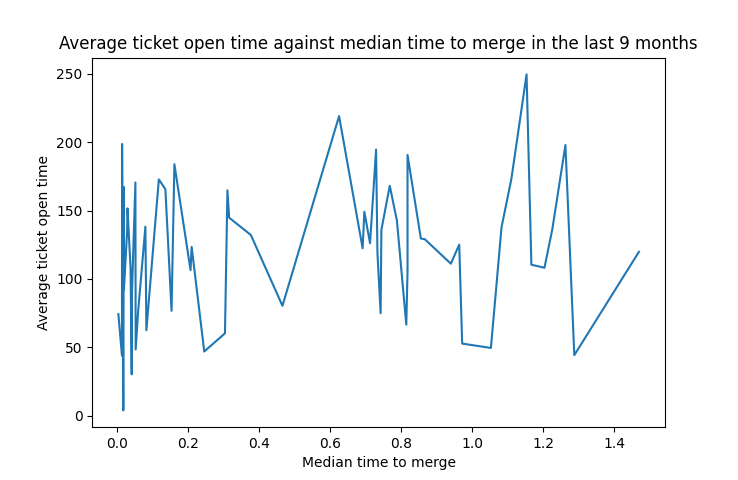} }}%
    \caption{\hyperref[list:item-9]{Average open time for tickets}}%
    \label{fig:average-ticket-time}%
\end{figure}

\begin{figure}[H]
    \centering
    \includegraphics[width=211pt]{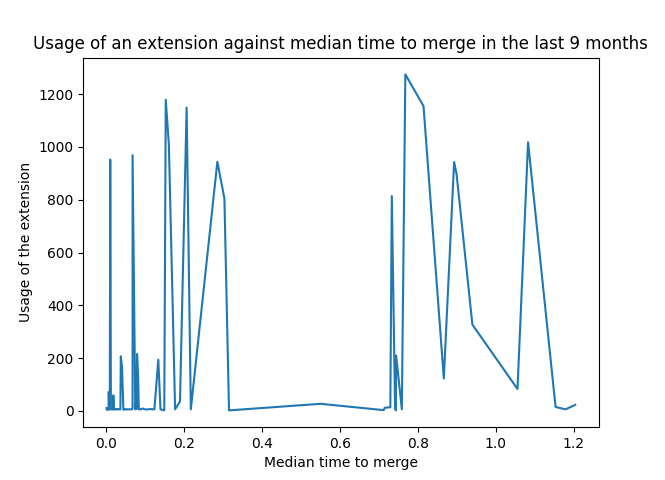}
    \caption{\hyperref[list:item-10]{Usage on wikis as reported by WikiApiary}}
    \label{fig:usage-9-m}
\end{figure}

\subsection{Pearsonr and p-value tests for graphed data}
\label{sec:Pearsonr}
\hspace{0.25cm}
\begin{center}
\begin{tabular}{@{}c c c c c@{}} 
 \hline
    \textbf{Variable} &
      \multicolumn{2}{c}{\textbf{Pearsonr coefficient}} &
      \multicolumn{2}{c}{\textbf{p-value}}  \\
      & {All} & {Last 9 months} & {All} & {Last 9 months} \\
\hline
 PHP test coverage percentage & N/A & 0.2066 & N/A & 0.1457 \\
 Number of submitted changes & 0.1384 & 0.2821 & \(9.218^{-5}\) & 0.0003 \\ 
 Number of merged changes & 0.1465 & 0.0903 & \(3.450^{-5}\) & 0.257 \\ 
 Number of authors of merged changes & 0.3077 & 0.4505 & \(7.361^{-19}\) & \(2.529^{-9}\) \\ 
 Number of code review votes & 0.1448 & 0.2635 & \(4.245^{-5}\) & 0.00079 \\ 
 Number of reviewers voting on code & 0.3193 & 0.5282 & \(3.017^{-20}\) & \(8.380^{-13}\) \\ 
 Average open time for tickets & 0.0625 & 0.1904 & 0.270 & 0.156 \\
 Usage on wikis as reported by WikiApiary & N/A & 0.3374 & N/A & 0.0005 \\ 
 \hline
\end{tabular}
\end{center}
\hspace{0.5cm}
\subsection{Tabled results}
\hspace{0.25cm}
\label{sec:steward-team}
\begin{center}
\begin{tabular}{@{}c c c c c c c@{}} 
 \hline
    \textbf{\hyperref[list:item-2]{Has a steward}} &
      \multicolumn{2}{c}{\textbf{Total}} &
      \multicolumn{2}{c}{\textbf{Average target variable}} &
      \multicolumn{2}{c}{\textbf{Median target variable}} \\
      & {All} & {Last 9 months} & {All} & {Last 9 months} & {All} & {Last 9 months} \\
\hline
 Yes & 122 & 77 & 1.1728 & 1.9989 & 0.7987 & 1.1104  \\
 No & 49 & 13 & 0.5311 & 1.6261 & 0.5125 & 0.7444  \\
 \hline
\end{tabular}
\end{center}

\label{sec:listed-maintainers}
\begin{center}
\begin{tabular}{@{}c c c c c c c@{}} 
 \hline
    \textbf{\hyperref[list:item-3]{Has a maintainer}} &
      \multicolumn{2}{c}{\textbf{Total}} &
      \multicolumn{2}{c}{\textbf{Average target variable}} &
      \multicolumn{2}{c}{\textbf{Median target variable}} \\
      & {All} & {Last 9 months} & {All} & {Last 9 months} & {All} & {Last 9 months} \\
\hline
 Yes & 107 & 39 & 0.9274 & 1.5408 & 0.6892 & 1.0826  \\
 No & 63 & 50 & 0.9482 & 2.2360 & 0.6507 & 0.9626  \\
 \hline
\end{tabular}
\end{center}
\hspace{0.5cm}
\subsection{t-statistic and p-values for tabled results}
\hspace{0.25cm}
\label{sec:t-statistic}
\begin{center}
\begin{tabular}{@{}c c c c c@{}} 
 \hline
    \textbf{Variable} &
      \multicolumn{2}{c}{\textbf{t-statistic}} &
      \multicolumn{2}{c}{\textbf{p-value}}  \\
      & {All} & {Last 9 months} & {All} & {Last 9 months} \\
\hline
 \hyperref[list:item-2]{Has a steward?} & 3.5442 & 0.4508 & 0.0005 & 0.1457 \\
 \hyperref[list:item-3]{Has at least one maintainer?} & -0.1368 & -1.1815 & 0.8913 & 0.2406 \\
 \hline
\end{tabular}
\end{center}
\hspace{0.25cm}
\section{Discussion}
\label{sec:discuss}

The p-value for \hyperref[list:item-1]{PHP test coverage} and \hyperref[list:item-9]{average open time of tickets} being greater than 0.05 indicates that these are uncorrelated to the target variable (median time to merge) and as such for these two the \(H_0\) hypothesis holds. There was not likely enough data to properly test the PHP test coverage as detailed in the results, so further research on other code quality metrics would be needed.

For graphed data (except the number of merged changes) data taken from the last 9 months gave a higher Pearsonr coefficient in general. The author thinks that this may be because the use of a larger period could lead to averaged-out periods of inactivity. As such, the author elects to prefer the data collected from the last 9 months unless the p-value for this 9-month data is too large (\(p > 0.05\)).

Interpreting the Pearsonr correlation coefficient is less standardised than interpreting the p-value \citep{Cohen2013}. As such the author proposes based on research from multiple sources that:
\begin{itemize}
    \item A value of 0.3 or below be considered to have no linear correlation and thus \(H_0\) holds.
    \item A value between 0.3 and 0.4 be considered a weak linear relationship, but not enough to reject \(H_0\).
    \item A value between 0.4 and 0.5 be considered a medium-strength linear relationship and it is then unclear whether the results support \(H_0\) or \(H_1\).
    \item A value above 0.5 be enough evidence to suggest that \(H_1\) holds and to also reject \(H_0\).
\end{itemize}

This means that the author's interpretations of the graphed results are:
\begin{itemize}
    \item For \hyperref[list:item-4]{variable numbers 4, 5, 7 and 10} \(H_0\) (the null hypothesis) holds.
    \item For \hyperref[list:item-6]{variable number 6}, the number of people who authored approved changes, the evidence does not support rejecting either \(H_0\) or \(H_1\), and as such further research is needed.
    \item For \hyperref[list:item-8]{variable number 8}, the number of reviewers reviewing code, there is evidence to reject \(H_0\).
\end{itemize}

As detailed at the top of the results section, the variables that were detailed in a table used a dataset with significantly less extensions. As such the author determines that these results should be given less weight for rejecting or accepting hypotheses.

Furthermore, for data from the last 9 months, the number of extensions reduces again. The author determines through this and the p-value tests that the size of the data is too small to draw solid conclusions from. However, using the data with no time limitations means that the median time to merge represents an average over a long period whereas \hyperref[list:item-2]{variables 2 and 3} only represent the current state of the extension. As such this reduces the usefulness of the results drawn from this data.

As such the author determines from the tabled results that:
\begin{itemize}
    \item For \hyperref[list:item-2]{variable number 2}, whether there is a listed steward, the results suggest that there is weak evidence to reject \(H_0\) and that further research would be required to get a better result.
    \item For \hyperref[list:item-3]{variable number 3}, whether there is at least one listed maintainer, the p-value suggests accepting \(H_0\).
\end{itemize}

This means the results find that the median time to merge increases when more reviewers vote on code. The results also suggest that an extension having a steward increases the median time for review.

While the accepted alternative hypotheses provide evidence for a correlation they do not necessarily find evidence for causation, as many factors other than the one variable may play into causing something \citep[pp. 25-26]{10.2307/43551404}. As such these results are limited to only providing evidence that these factors may be linked, but does not prove that this is the sole or major cause of a change in the target variable.

These results only measured data for extensions for the MediaWiki project. There are other parts of the MediaWiki project, including the core installation and other repositories not under the scope of extensions. As such applying these results without further research to the wider open-source project landscape, or even to another part of the MediaWiki project, may be difficult to achieve without further research.

\section{Conclusion \& Future Work}
\label{sec:conc}

The author has found that more reviewers voting on code decreases the speed of code review and also that having a steward may decrease the speed of code review.

The author notes that the correlation in these results may be due to other underlying factors. For example:
\begin{itemize}
    \item An extension having a steward may be due to it being more important, which could lead to more changes that could in turn backlog the review process (leading to a decrease in code review speed).
    \item The number of reviewers reviewing code may also be due to the extension being more important, which could naturally increase the number of people willing to review code for the repository.
\end{itemize}

Future work could include extending the analysis to other variables for each extension and choosing data from multiple specific periods of time including periods that start and end in the past. However, doing further analysis may require sampling the extensions to find a good group of extensions that accurately represent the whole to make analysis and data collection quicker and easier.

If readers want to extend this work to other areas of the MediaWiki project or more generally to open-source development further data collection is likely needed to build upon these results so that they can be validated to apply more generally.

The author intends to carry out research on recommending reviewers automatically for patches in the context of MediaWiki for their thesis paper next semester. As part of this the author is thinking of working out whether the automatic recommendation improves the speed of code review.

\section{Reflective Analysis}
The author of this paper believes that he did not give himself enough time to work on the project so that it would be ready for the original deadline. This assessment and a group project for another course were due in at the same time. While my group did make good progress to plan getting in both assignments on time, the group work took much longer than expected. As such it was submitted a few hours before the deadline, which was much longer after we had planned to submit. This meant this assessment was pushed back. The extension allowed me to get this report done to a standard I was happier with.

If I was to do this assessment again I would aim to build in more buffer time so that I could respond to changes in plan, such other assessments taking longer than expected. I would have also generated the results at an earlier date so that I could more effectively review the results and decide if I needed to choose different hypotheses so that I had enough to discuss and draw conclusions from in the report.

\newpage

\bibliography{myrefs}

\end{document}